\documentclass[10pt, conference]{IEEEtran}
\usepackage[utf8]{inputenc}
\usepackage[detect-all]{siunitx}
\usepackage{array,tabularx,calc}
\usepackage{mathtools}
\usepackage{amsfonts}
\usepackage{graphicx}
\usepackage{amssymb}
\usepackage{amsfonts}
\usepackage{pbox}
\usepackage{url}
\usepackage[]{algorithm, algpseudocode}
\usepackage{tikz}

\DeclareSIUnit{\belmilliwatt}{Bm}
\DeclareSIUnit{\dBm}{\deci\belmilliwatt}
\DeclareSIUnit{\belisotropic}{Bi}
\DeclareSIUnit{\dBm}{\deci\belisotropic}
\DeclareSIUnit{\bit}{bit}

\ifCLASSOPTIONcompsoc	\usepackage[caption=false,font=normalsize,labelfont=sf,textfont=sf]{subfig}
\else					\usepackage[caption=false,font=footnotesize]{subfig}
\fi

\makeatletter
\usepackage[]{algorithm, algpseudocode}
\makeatletter
\renewcommand{\ALG@beginalgorithmic}{\small}

\usepackage[capitalise]{cleveref}	
    \crefformat{equation}{(#2#1#3)}	

\usepackage[noadjust]{cite}

\begin{document}

\usetikzlibrary{arrows}
\usetikzlibrary{shapes}
\newcommand{\mymk}[1]{%
	\tikz[baseline=(char.base)]\node[anchor=south west, draw,rectangle, rounded corners, inner sep=0.1pt, minimum size=3.5mm,
	text height=2mm](char){\ensuremath{#1}} ;}

\newcommand*\circled[1]{\tikz[baseline=(char.base)]{
	\node[shape=circle,draw,inner sep=0.1pt] (char) {#1};}}

\title{Joint Traffic and Obstacle-aware UAV Positioning Algorithm for Aerial Networks }

\author{\IEEEauthorblockN{Kamran Shafafi, André Coelho, Rui Campos, Manuel Ricardo}
	\IEEEauthorblockA{INESC TEC and Faculdade de Engenharia, Universidade do Porto, Portugal\\
		\{kamran.shafafi, andre.f.coelho, rui.l.campos, manuel.ricardo\}@inesctec.pt}}

\maketitle

\begin{abstract}

Unmanned Aerial Vehicles (UAVs) are increasingly used as cost-effective and flexible Wi-Fi Access Points (APs) and cellular Base Stations (BSs) to enhance Quality of Service (QoS). In disaster management scenarios, UAV-based networks provide on-demand wireless connectivity when traditional infrastructures fail. In obstacle-rich environments like urban areas, reliable high-capacity communications links depend on Line-of-Sight (LoS) availability, especially at higher frequencies. Positioning UAVs to consider obstacles and enable LoS communications represents a promising solution that requires further exploration and development.\looseness=-1 

The main contribution of this paper is the Traffic- and Obstacle-aware UAV Positioning Algorithm (TOPA). TOPA takes into account the users' traffic demand and the need for LoS between the UAV and the ground users in the presence of obstacles. The network performance achieved when using TOPA was evaluated through ns-3 simulations. The results show up to 100\% improvement in the aggregate throughput without compromising fairness .\looseness=-1
  
\end{abstract}


\section{Introduction}
In recent years, Unmanned Aerial Vehicles (UAVs) gained attention for deploying versatile Wi-Fi Access Points (APs) and cellular Base Stations (BSs) due to low cost and flexible deployment \cite{rs11121443, 8660516}. UAV-based networks offer unique advantages, especially in disaster management scenarios such as wildfires, earthquakes, floods, cyber, and terrorist attacks, where traditional infrastructures may fail \cite{8875210}. The need to enhance wireless network coverage for ubiquitous Internet access is crucial in today's digital society, enabling various online services and applications such as ultra-high definition videos, online games, augmented reality, disaster safety, and event facilitation.\looseness=-1 

In this context, Flying Networks (FNs), formed by UAVs, balloons, and airships, have emerged as an effective and flexible solution to provide wireless communications anywhere and anytime \cite{6735774, 8438896}. By using the capabilities of UAVs, FNs can outperform traditional terrestrial networks and provide communications in remote areas, disaster-stricken regions, and extreme environments. However, the successful deployment and operation of FNs introduce their own challenges. One of the key challenges is to ensure Line-of-Sight (LoS) availability at high frequencies, in which the radio signal propagation can be easily obstructed by obstacles, such as buildings, vegetation, and terrain irregularities. These obstructions can significantly limit the coverage range and reliability of FNs.\looseness=-1

\begin{figure}
	\centering
	\includegraphics[width=1\linewidth]{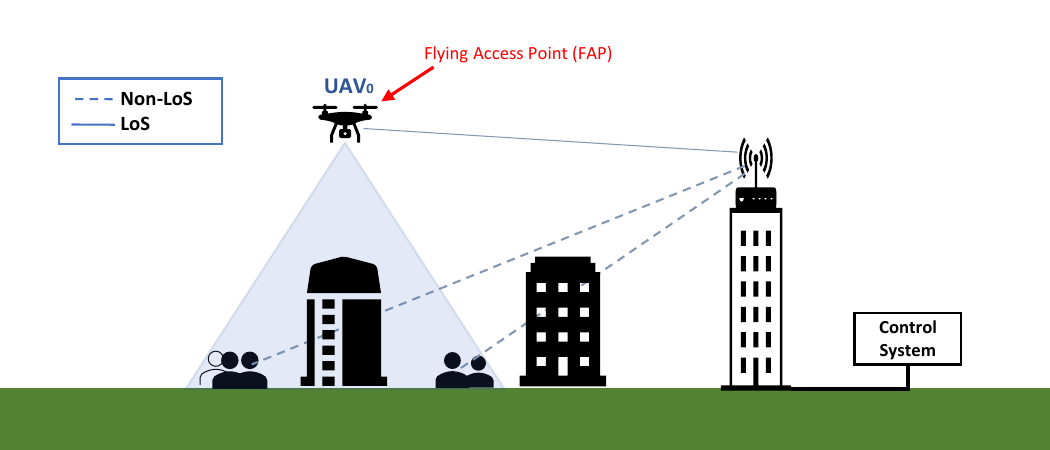}
	\caption{Flying network consisting of UAV positioned to provide LoS wireless connectivity. The users are served by the UAV, which is connected to the LTE Base Station (BS). }
	\label{fig1}
\end{figure}

To overcome this challenge, as depicted in \cref{fig1}, the solution lies in exploring positioning approaches for UAVs to ensure LoS and optimize network performance. By optimally positioning UAVs, it becomes possible to establish LoS communications links between UAVs and User Equipment (UE), ensuring maximum network performance even in environments with obstacles. However, existing solutions for improving wireless network coverage typically overlook the presence of obstacles in the environment or do not fully address the challenges associated with positioning the FNs in such scenarios \cite{ALMEIDA2021102525, 9448966, 8589098, 9217183, 9755063}.\looseness=-1

The main contribution of this paper is the Traffic- and Obstacle-aware UAV Positioning Algorithm (TOPA). TOPA takes into account the users' traffic demand and the need for LoS between the UAV and the ground users in the presence of obstacles. TOPA is designed to address the challenges posed by environments with obstacles, ensuring LoS connectivity and meeting the traffic demand of each user. By integrating traffic awareness into the positioning algorithm, TOPA establishes high throughput communications links with high enough capacity to accommodate the user traffic demand. Simulation results using Network Simulator (ns-3) validate the effectiveness of TOPA in improving network performance and ensuring LoS, particularly in terms of aggregate throughput  between the ground users and the UAV.\looseness=-1
 
The rest of this paper is organized as follows. \cref{sec:soa} provides an overview of the state-of-the-art positioning approaches in FNs. \cref{sec:system_model} introduces the system model. \cref{sec:problem_formulation} formulates the problem. \cref{Traffic- and Obstacle-aware Positioning Algorithm} presents the TOPA, including its underlying rationale and a demonstration for a simplified scenario. \cref{sec:performance_evaluation} focuses on the performance evaluation, including the simulation setup, scenarios, performance metrics, and simulation results. Finally, \cref{sec:conclusions} point out the main conclusions and outlines potential future work directions.\looseness=-1

\section{State of the Art~\label{sec:soa}}
In this section, we provide a review of the existing solutions in the literature regarding the positioning problem of FNs and highlight their major limitations.\looseness=-1

The research works mentioned in this paragraph primarily focus on obstacle-free environments, neglecting the impact of obstacles on LoS. In \cite{ALMEIDA2021102525}, a traffic-aware positioning solution was proposed to meet user traffic demand. The UAV's position is adjusted based on the estimated user location on the ground. Similarly, \cite{9448966} presents a heuristic positioning solution for a set of UAVs functioning as a backhaul network. \cite{8589098} introduces a centralized routing solution for a flying multi-hop network, aiming to establish high-capacity paths between UAVs. In \cite{9217183}, a proactive queue management solution for FNs with controlled topology was proposed, in order to define the UAV queue size over time to maximize throughput while ensuring stochastic delay guarantees. \cite{7486987} proposes an optimal 3D deployment strategy for multiple UAVs to maximize downlink coverage performance while minimizing transmission power. \cite{7762053} focuses on utilizing UAVs as flying Base Stations (BSs), optimizing the number of UAVs required to provide coverage for the users. However, these works can be enhanced by considering the presence of obstacles.\looseness=-1

There are other state-of-the-art works that specifically address the issue of potential blockages faced by UEs when encountering obstacles. In \cite{9681949}, the authors propose an end-to-end system that tackles the blockage prediction problem for UEs by leveraging point cloud processing and deep learning techniques. Another solution presented in \cite{9473651} proactively predicts dynamic link blockages between a 5G BS and the user, utilizing video-based approaches. Furthermore, \cite{9464922} demonstrates how computer vision techniques enable look-ahead prediction in scenarios involving millimeter-wave channel blockages, enabling proactive measures before the actual blockage occurs. However, these works primarily focus on providing wireless paths through terrestrial BSs and overlook the positioning of UAVs in FNs.\looseness=-1

\section{System Model~\label{sec:system_model}}
The system model consists of: i) a UAV as a Flying AP (FAP) to provide wireless communications to the users; ii) $N-1$ UEs distributed in the venue and waiting to be served by FAP; iii) a building as obstacle deployed in the scenario; iv) a Control System (CS) that can be deployed anywhere in the cloud, edge, or in the UAV, as depicted in \cref{fig1}. The scenario aims to take advantage of short-range, high-directional air-to-ground radio links. In more crowded scenarios, an effective approach is to increase the number of UAVs and create isolated scenarios, with each UAV assigned to a specific set of UEs. This allows for better resource allocation and improved network efficiency. In this paper, we focus on the single-UAV scenario and leave the multi-UAV scenarios for future work.\looseness=-1

The CS is responsible for: 1) collecting the coordinates of the UEs on the ground by running a Position Recognition Algorithm (PRA) such as the one presented in \cite{deng2018wifi} or similar algorithms to determine the coordinates of the UEs -- this information is critical for positioning the UAV; 2) updating the coordinates of the obstacles using the map of the venue -- Herein, only buildings are considered as obstacles, leaving the consideration of moving obstacles, like cars, for future research; 3) determining the updated position of the UAV, taking into account the provided information, to establish LoS with all UEs and accommodate their traffic demand. The resulting optimal positions are sent by the CS to the UAV, which positions itself accordingly.\looseness=-1

\section{Problem Formulation~\label{sec:problem_formulation}}
The flying network is represented by a directed graph, $G(t_k)=(U, L(t_k))$, where $t_k = k \times \Delta t, k \in \mathbb{N}_0$ and $\Delta t \in \mathbb{R}$, represents the time instant to update the UEs position by PRA. $U$ comprises $N$ nodes forming the network, including all the UEs and the UAV. The links between the UEs and the UAV at time $t_k$ are represented by the set $L(t_k)$.\looseness=-1 

Let us consider the $UE_i$, where $i\in \{1, ..., N-1\}$, transmits a traffic flow of bitrate $T_i(t_k)$ \SI{}{bit/s} during time slot $t_k$ to the $UAV$. The flow bitrate, $T_i(t_k)$, in \SI{}{bit/s} requires a minimum capacity $C_{i}(t_k)$, in bit/s, for the bidirectional wireless link between $UE_i$ and the $UAV$ at time $t_k$. The flow is received at the $UAV$ from $UE_i$ with a bitrate of $R_i(t_k)$ bit/s. The maximum channel capacity is $C^{MAX}$ \SI{}{bit/s}. We assume the wireless medium is shared, the $UAV$ has the ability to listen to any $UE_i$, and the Carrier Sense Multiple Access with Collision Avoidance (CSMA/CA) mechanism is employed for Medium Access Control (MAC). \looseness=-1 

Our objective is to determine, at any given time $t_k$, the position of the $UAV$, represented as $P= (x, y, z)$, which ensures LoS connectivity with all UEs. Additionally, we aim to determine the transmission power $P_T$ of the UEs which we assume will be equal for all of them. This computation takes into account the following rationale: the maximum transmission power $P_T^{MAX}$ allowed for the wireless technology used ensures that the capacity $C_{i}(t_k)$ is sufficient to accommodate the bitrate $T_i(t_k)$ while minimizing the $C(t_k) = \sum_{i=1}^{N-1}C_{i}(t_k)$, where $C(t_k)$ is the overall network capacity. By minimizing the $C(t_k)$, we can reduce the required transmission power, leading to less interference between nodes and lower energy consumption. In our scenario, $\theta_{i1}$ is the angle between the ground and the line that passes through $UE_i$ and the corner of the building closest to $UE_i$ (solid lines in \cref{fig2}) and $\theta_{i2}$  is the angle between the ground and the line that passes through $UE_i$ and the $UAV$ (dashed line in \cref{fig2}).\looseness=-1

In simple terms, the problem can be formulated as defined in \cref{eq:objective-function1}.\looseness=-1

\begin{small}
\begin{subequations}\label{objectives}    
    \begin{alignat}{8}
            & \!\underset{P_T, (x, y, z)}{\textrm{minimize}} &   & C(t_k)=\sum_{i=1}^{N-1}C_{i}(t_k)\label{eq:objective-function1}\\              
		  & \text{subject to:} & &~0 \leq P_T \leq P_T^{MAX}\label{eq:constraint1}\\
		  &     &     &~C(t_k) \leq C^{MAX}\label{eq:constraint2}\\
            &     &     &~0 < T_i(t_k) \leq C_{i}(t_k)\hspace{1.2cm} i \in \{1,...,N-1\}\label{eq:constraint3}\\
		   &     &     &~0 \leq (x, y, z) \leq (x^{MAX}, y^{MAX}, z^{MAX})\notag\\
           &     &     &\hspace{4cm}i \in \{1,...,N-1\}\\     
            &    & & \theta_{i1} \leq \theta_{i2},\hspace{2.6cm}i \in \{1, ..., N-1\}\label{eq:constraint4}     
	\end{alignat}    
\end{subequations}
\end{small}\vspace{0.1cm}

Where, $(x^{MAX}, y^{MAX}, z^{MAX})$ is the maximum coordinate allowed, in which the $UAV$ can be deployed to accommodate the $T_i(t_k)$.\looseness=-1

\section{Traffic- and Obstacle-aware Positioning Algorithm\label{Traffic- and Obstacle-aware Positioning Algorithm}}
\begin{figure}
	\centering
		\includegraphics[width=\linewidth]{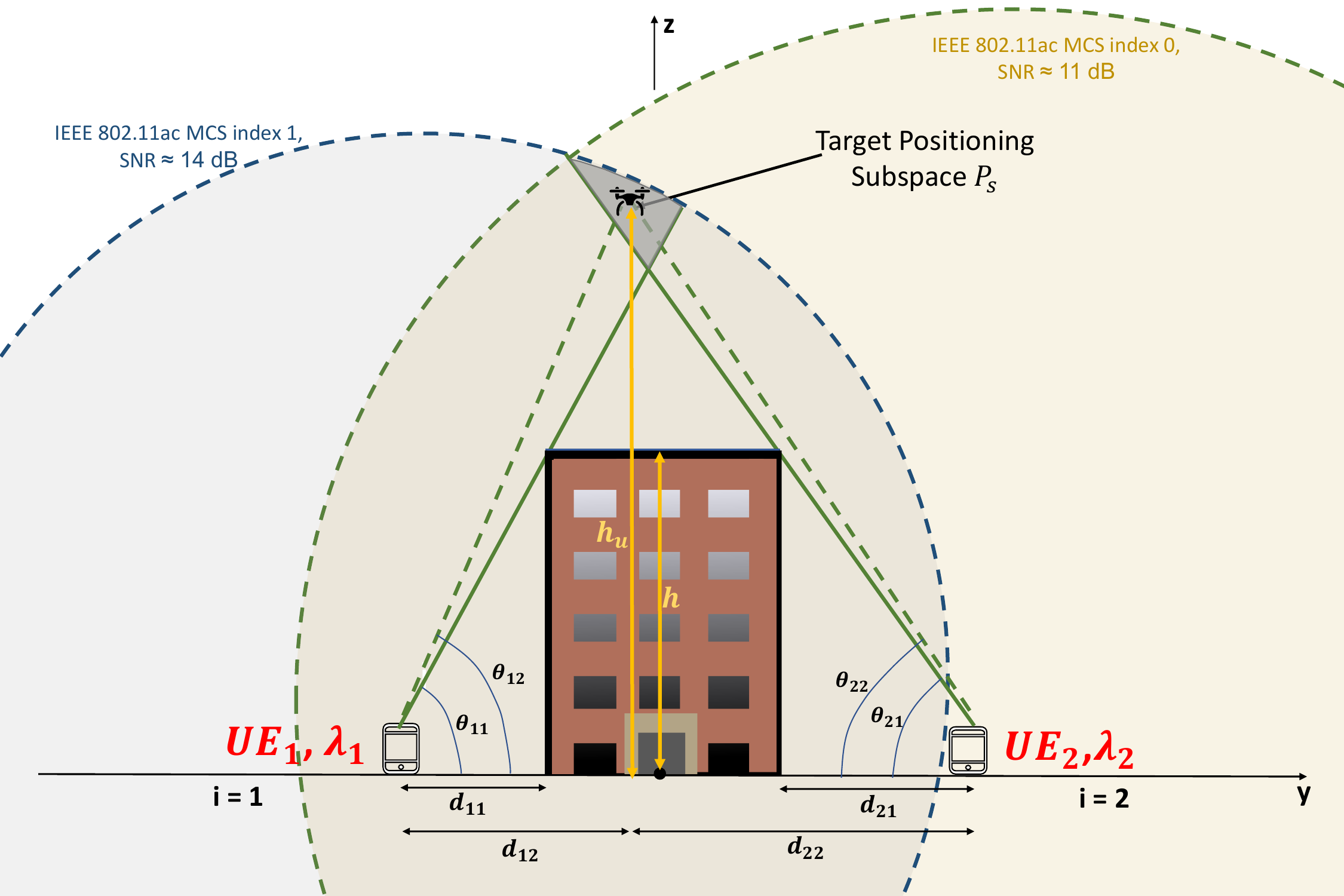}
	\caption{Two-dimensional (2D) representation of the target positioning subspace ($P_s$), which is determined by the intersection of LoS constraints derived from the obstacle and the traffic demand spheres centered at each UE.}
	\label{fig2}
\end{figure}

The target positioning subspace ($P_s$) is identified as the area for positioning the $UAV$, as depicted in \cref{fig2}. All the points in this subspace are in LoS with all UEs and satisfy their traffic demands. Among these points, $P= (x, y, z)$, the optimal position, minimizes both the sumrates and the transmitted energy.\looseness=-1

First, TOPA assumes the environment is free of obstacles. This stage aims at ensuring that the wireless link established between $UE_i$ and the $UAV$ has a minimum Signal-to-Noise Ratio (SNR), denoted as $SNR_{i}$, which enables the utilization of an appropriate Modulation and Coding Scheme (MCS) index, denoted as $MCS_i$. The selected $MCS_i$ should be capable of transporting $T_i$~\SI{}{bit/s}. By guaranteeing the conditions for $MCS_i$, we can ensure that the network has the capacity to accommodate $T_i$, thereby maximizing the amount of data received at the UAV, $R(t_k)$. Additionally, the selection of $MCS_i$ imposes a minimum $SNR_i$, assuming a constant noise power $P_N$. Given the known transmission power $P_T$, we can calculate the maximum distance $d_{\text{max}_i}$ between $UE_i$ and the $UAV$ using the Free-space path loss model defined in Eq. \ref{eq:friis-propagation-model} in \SI{}{\decibel}, where $f_i$ represents the carrier frequency, in Hz, and $c$ corresponds to the speed of light in vacuum.\looseness=-1 

\begin{small}
\begin{equation}
	\begin{aligned}
	SNR_{i} \smash{=} P_{T} \smash{-} 20\log_{10}(d_{max_i}) \smash{-} 20\log_{10}(f_i) \smash{-} 20\log_{10} \bigl(\begin{smallmatrix}
	\frac{4\smash{\times}\pi}{c}
	\end{smallmatrix}\bigr) 
	\smash{-}P_N
	\end{aligned}
	\label{eq:friis-propagation-model}
\end{equation}
\end{small}

Taking into account $N-1$ UEs, the target positioning subspace $P_s$ for placing the $UAV$ is determined by the intersection of the spheres centered at each $UE_i$. The maximum distance $d_{\text{max}_i}$ represents the radius of the sphere centered at $UE_i$, indicating the range within which the $UAV$ should be positioned. By solving Eq. \cref{eq:placement-equations-system}, derived from Eq. \cref{eq:friis-propagation-model} for each $UE_i$ at $(x_i, y_i, z_i)$, TOPA calculates the $UAV$ coordinates, $(x, y, z)$.\looseness=-1 

\begin{equation} \label{eq:placement-equations-system}
\begin{aligned}
    & (x - x_i)^2 + (y - y_i)^2 + (z - z_i)^2 \leqslant \left(10^{\frac{K + P_T - SNR_i}{20}}\right)^2 \\
    & K = -20\log_{10}(f_i) - 20\log_{10}\left(\frac{4\times\pi}{c}\right) - (P_N)
\end{aligned}
\end{equation}\vspace{0.1cm}

Secondly, TOPA focuses on enforcing constraints to ensure LoS connectivity between the UEs and the $UAV$, resulting in the positioning subspace $P_s$. As depicted in \cref{fig2}, for establishment LoS, $\theta_{i1}\smash{=} arctan\left(\frac{d_{i1}}{h}\right)$ $\leqslant$ $\theta_{i2}\smash{=} arctan\left(\frac{d_{i2}}{h_u}\right)$.\looseness=-1

$P_s$ is obtained by considering the volume defined by the intersection of the spheres centered at each $UE_i$ and the lines defining the angle $\theta_{i2}$ with the ground. $P_s$ is represented by the gray area in \cref{fig2}. TOPA assumes symmetry links and constant transmission power for all UEs and the UAV.\looseness=-1   

\begin{algorithm}
	\renewcommand{\thealgorithm}{A}
	\caption{-- TOPA}
	\label{algorithm}
	\begin{algorithmic}[1]
		\State $P_T = 0$ \Comment{\SI{0}{dBm} TX power}
        \While {$P_T \leq P_T^{MAX}$} \Comment{Allowed TX power}               
		    \State $P_{T_i} = P_T, i\in \{1, ..., N-1\}$ \Comment{UEs' TX power}
            \If{$(x, y, z) \leq (x^{MAX}, y^{MAX}, z^{MAX})$}   
                \If{$\theta_{i1} \leq \theta_{i2}$}  
		            \State Calculate $(x_0, y_0, z_0)$ 
		            \If{$(x, y, z) \neq \oslash$} \Comment{$(x, y, z) \in P_s$}
		                \State \textbf{return} $P_T, (x, y, z)$ 
		            \Else
		                \State $P_T = P_T + 1$ \Comment{Increase TX power by \SI{1}{\deci\belmilliwatt}}
		            \EndIf        
                \EndIf             
            \EndIf
        \EndWhile
    \end{algorithmic}
\end{algorithm}

We provide an illustrative example of the execution of TOPA in the scenario depicted in \cref{fig2}. TOPA can be used in scenarios consisting of $N-1$ UEs, one UAV, and one obstacle. For illustration purposes, we assume the obstacle has dimensions of (10, 10, 20) meters, respectively length, width, and height. We also consider two UEs, $UE_1$ and $UE_2$, which are located at coordinates (0, -15, 1) and (0, 20, 1), respectively; we consider an altitude of 1~m for each UE to represent UEs carried by people. We consider the utilization of the IEEE 802.11ac standard with one spatial stream, a  Guard Interval (GI) of \SI{800}{\nano \second}, and channel bandwidth of \SI{160}{\mega\hertz} (channel 50 at 5250 MHz). We assume that the demanded capacity for $UE_1$ is \SI{117}{Mbit/s}, associated with the IEEE 802.11ac MCS index 1, and the demanded capacity for $UE_2$ is \SI{58.5}{Mbit/s}, associated with the IEEE 802.11ac MCS index 0 \cite{MCSwebsite}. These target MCS indexes are considered to illustrate the execution of \cref{algorithm} where $UE_1$ has a traffic demand two times higher than $UE_2$.\looseness=-1

Based on the minimum Received Signal Strength Indicator (RSSI) values proposed in \cite{MCSwebsite} and a noise floor power of \SI{-85}{dBm}, the minimum target SNR values in \SI{}{\decibel} are \SI{14}{\deci\bel} for $UE_1$ and \SI{11}{\deci\bel} for $UE_2$. By solving the system of equations \cref{eq:placement-equations-system} derived from the Friis propagation model \cref{eq:friis-propagation-model}, consideration the LoS constraint by \cref{eq:constraint4}, and replacing $k$ from \cref{k}, we determine the optimal position for the UAV and its transmission power using the optimization solver $GEKKO$. The optimal position is $(x, y, z) \approx (0.0, -1.48, 29.44)$, with a transmission power $P_{T}$ of \SI{6} {\deci\belmilliwatt}. The parameter $P_T$ in \cref{eq:placement-equations-system} serves as a fine-tuning parameter and is initially set to \SI{0}{\deci\belmilliwatt}. The algorithm proceeds by incrementing the value of $P_T$ in steps of \SI{1}{\deci\belmilliwatt} until a valid solution is achieved. If no solution is found within the specified range, the execution of TOPA is halted. TOPA aims at reducing the distances of wireless links while ensuring that the transmission power $P_T$ remains within the prescribed maximum value $P_T^\textrm{MAX}$.\looseness=-1

\begin{small}
\begin{equation}
	\begin{aligned}
	& K \smash{=}\smash{-}20\log_{10}(5250\smash{\times}10^6) - 20\log_{10}\left(\frac{4\times\pi}{3\smash{\times}10^8}\right) - (-85)
	\end{aligned}
	\label{k}
\end{equation}
\end{small}

\section{Performance Evaluation~\label{sec:performance_evaluation}}
This section presents the simulation setup, an overview of the simulation scenarios used, and an analysis of the simulation results obtained for TOPA using ns-3.\looseness=-1

\subsection{Simulation Setup~\label{sec:simulation_setup}}

We developed a framework in ns-3 to evaluate the performance of the proposed algorithm. In this framework, we deployed the building as a static obstacle using $Building Model$ and the users in a fixed position with $ConstantPositionMobilityModel$. A Network Interface Card (NIC) was configured on UAV in Ad Hoc mode, using the IEEE 802.11ac standard in channel 50, with a channel bandwidth of 160 MHz, and GI of 800 ns. One spatial stream was used for all links between UEs and the UAV. UDP traffic was generated by the $OnOffApplicationModule$ on each UE, while a UDP sink receiver was installed on the UAV. The framework evaluates the scenarios in two phases.\looseness=-1 

In the first phase, based on the coordinates of the UEs and the position of the UAV, the framework checks the channel condition regarding the establishment of LoS or NLoS links between the UAV and the UEs considering the obstacles. The $ItuR1411LosPropagationLossModel$ is applied when the UAV has LoS with UEs. Otherwise, the $ItuR1411NlosOverRooftopPropagationLossModel$ is applied.\looseness=-1

In the second phase, the performance of the network is evaluated by generating UDP traffic with a constant packet size of 1024 bytes. The data rate is adjusted to meet each UEs' traffic demand, ensuring compatibility and efficient data transmission. The $IdealWifiManager$ auto rate mechanism was utilized. The simulation is conducted over a duration of 100 seconds, measuring the aggregate throughput on the UAV. The evaluation interval is set to $1s$ to capture the average aggregate throughput over time.\looseness=-1

\subsection{Simulation Scenarios~\label{sec:simulation_scenarios}}
\begin{figure}
	\centering
		\includegraphics[width=\linewidth]{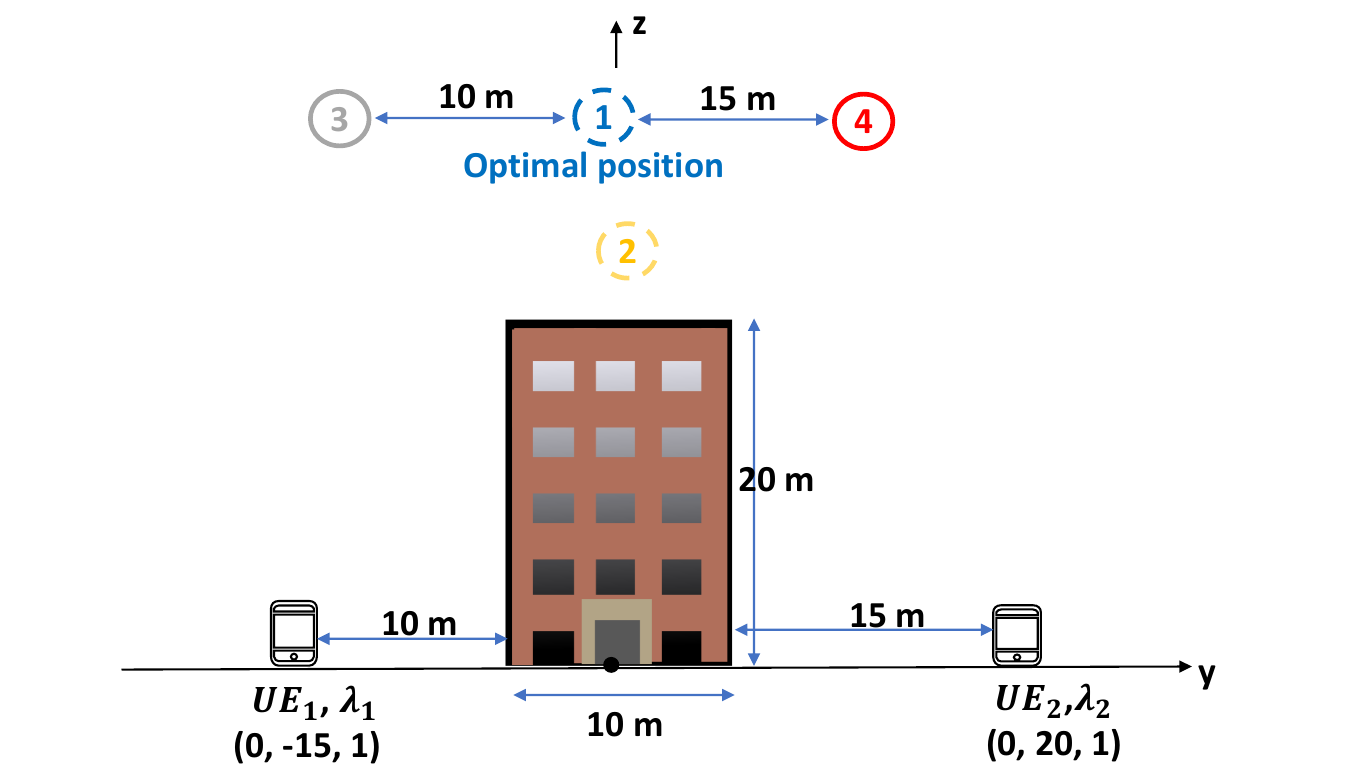}
	\caption{Scenario A -- two UEs and four possible positions. Position~1 is the optimal solution defined by TOPA, which allows LoS with all UEs, while Position 2 is the baseline, which is located five meters above the middle of the rooftop of the building.}
   \label{fig3}
\end{figure}

We have considered two scenarios to evaluate the performance of TOPA. For scenario A, depicted in \cref{fig3}, the optimal position achieved by TOPA (Position 1) provides LoS connectivity with both UEs over the rooftop. In order to assess the performance of TOPA, we consider a total of 4 positions, including the point, which is randomly located above the middle of the rooftop of the building (Position~2). Position~3 and position~4 are located, respectively, in LoS with $UE_1$ and $UE_2$. To make the evaluation of TOPA more realistic and suitable for heterogeneous scenarios, we introduce varying traffic demands for the UEs. Specifically, the traffic demand of $UE_1$ is set to be twice the traffic demand of $UE_2$. This difference in traffic demands allows us to assess the performance of TOPA under varying load conditions and capture the impact of asymmetrical traffic requirements ($\lambda_1 \smash{=} 2 \times \lambda_2$).\looseness=-1

\begin{figure}
	\centering
		\includegraphics[width=\linewidth]{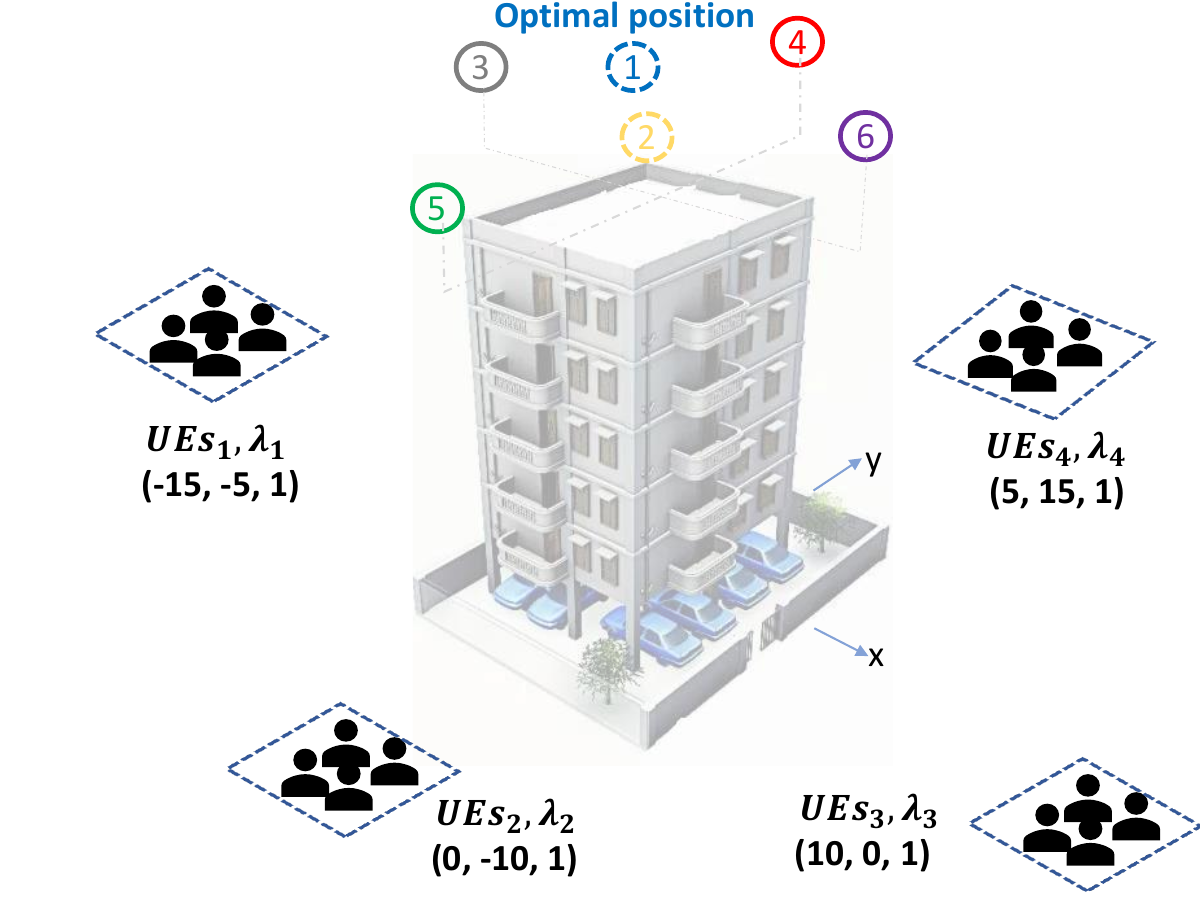}
	\caption{Scenario B -- four groups of UEs located on different sides of the building. Each group is represented by a square with dimensions of $1m \times 1m$ and associated traffic demand that is presented by $\lambda_i, i\in \{1, ..., N-1\}$.}
	\label{fig4}
\end{figure}

Scenario B, depicted in Fig. 4, aims at demonstrating the versatility and effectiveness of TOPA in crowded scenarios with multiple UEs. This scenario serves as a reference to assess the performance and capabilities of TOPA in challenging and densely populated environments. To simulate a diverse and realistic environment, we have divided the UEs into four groups located on different sides of the building; without loss of generality, the groups are placed in front of each facade of the building. Each group is represented by a square with dimensions of $1m \times 1m$, and the center of each square corresponds to the location of the respective group. This simplification allows us to capture the spatial distribution of UEs in the scenario while maintaining a manageable representation of their positions. Each group has its own traffic demand, which is presented by $\lambda_i,  i\in \{1, ..., N-1\}$. The scenario is evaluated in a total of 6 positions to assess the performance of TOPA. Position 1 represents the optimal placement achieved by TOPA, which maximizes network throughput. The remaining positions are situated on different sides of the building and (Position~2) is randomly located above the middle of the rooftop of the building. Based on the same rationale, three different traffic demand combinations are considered: i) $\lambda_1 ~\smash{=} ~\lambda_2 ~\smash{=} ~\lambda_3 ~\smash{=} ~\lambda_4$; ii) $\lambda_1 ~\smash{=} ~\lambda_2 ~\smash{=} ~2 \times \lambda_3 ~\smash{=} ~2 \times \lambda_4$; iii) $\lambda_1 ~\smash{=} ~2 \times \lambda_2 ~\smash{=} ~4 \times \lambda_3 ~\smash{=} ~8 \times \lambda_4$. \looseness=-1 

\subsection{Simulation Results~\label{sec:simulation_Results}}

In this section, we present the simulation results obtained from a 100-second simulation run for various traffic demand combinations, as described in \cref{sec:simulation_scenarios}. The average aggregate throughput received by the UAV is measured over each one-second simulation time. The results are depicted using the complementary cumulative distribution function (CCDF) to illustrate the distribution of the aggregate throughput across different scenarios.\looseness=-1

\begin{figure}
	\centering
		\includegraphics[width=\linewidth]{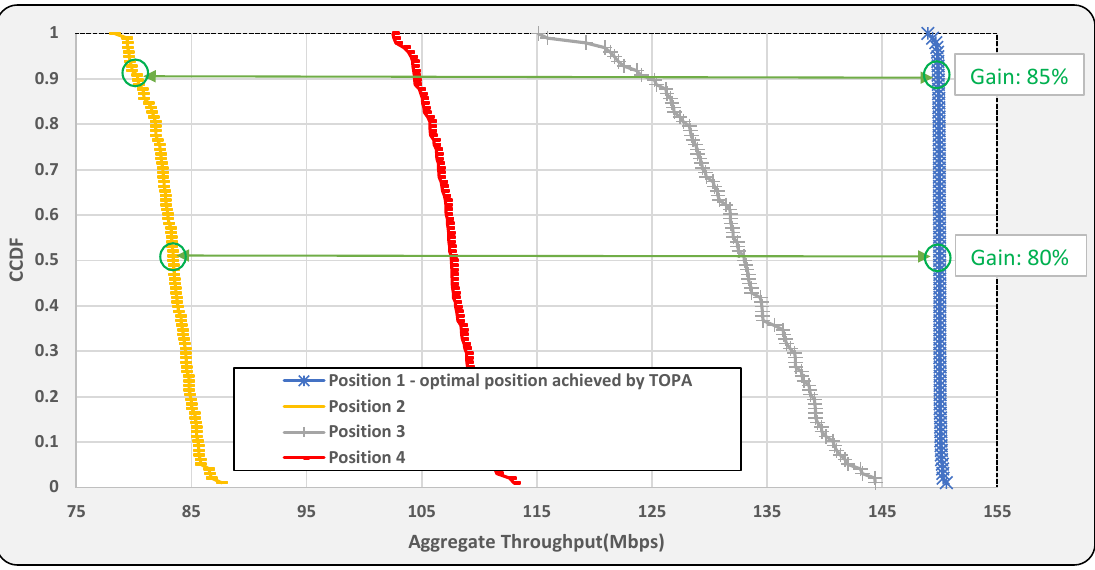}
	\caption{Scenario A - Aggregate throughput measured on UAV, where $\lambda_1 \smash{=} 2 \times \lambda_2$. }
	\label{fig5}
\end{figure}

\begin{figure}
	\centering
		\includegraphics[width=\linewidth]{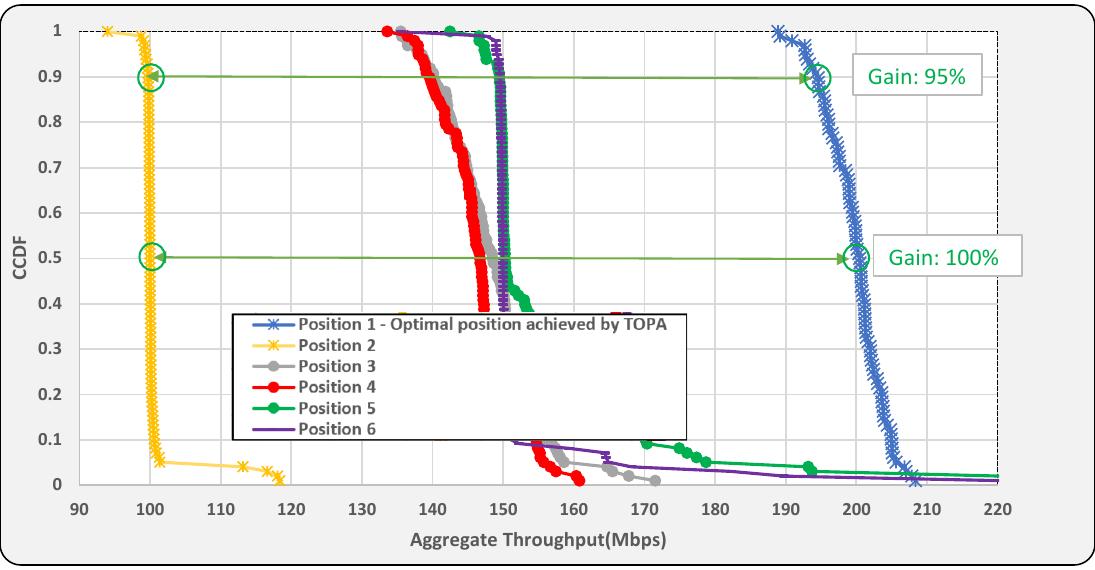}
	\caption{Scenario B - Aggregate throughput measured on UAV, where $\lambda_1 ~\smash{=} ~\lambda_2 ~\smash{=} ~\lambda_3 ~\smash{=} ~\lambda_4$.}
	\label{fig6}
\end{figure}

In Scenario A, shown in \cref{fig5}, placing the UAV in the optimal position (Position 1 in \cref{fig3}) leads to significant improvement in aggregate throughput. Specifically, compared to Position 2, the aggregate throughput is enhanced up to 85\% for the 90th percentile and up to 80\% for the 50th percentile (median). In positions 3 and position 4 the $UAV$ is not in LoS with all UEs. Additionally, considering the varying traffic demand, the aggregate throughput on the UAV differs across these positions.\looseness=-1 

\begin{figure}
	\centering
		\includegraphics[width=\linewidth]{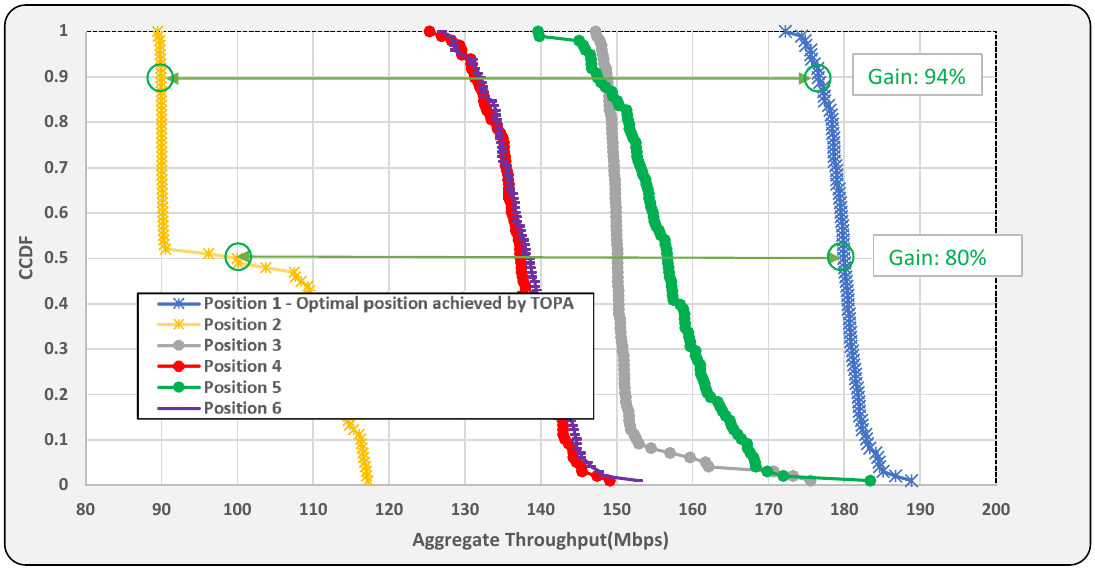}
	\caption{Scenario B - Aggregate throughput measured on UAV, where $\lambda_1 ~\smash{=} ~\lambda_2 ~\smash{=} ~2 \times \lambda_3 ~\smash{=} ~2 \times \lambda_4$.}
	\label{fig7}
\end{figure}

\begin{figure}
	\centering
		\includegraphics[width=\linewidth]{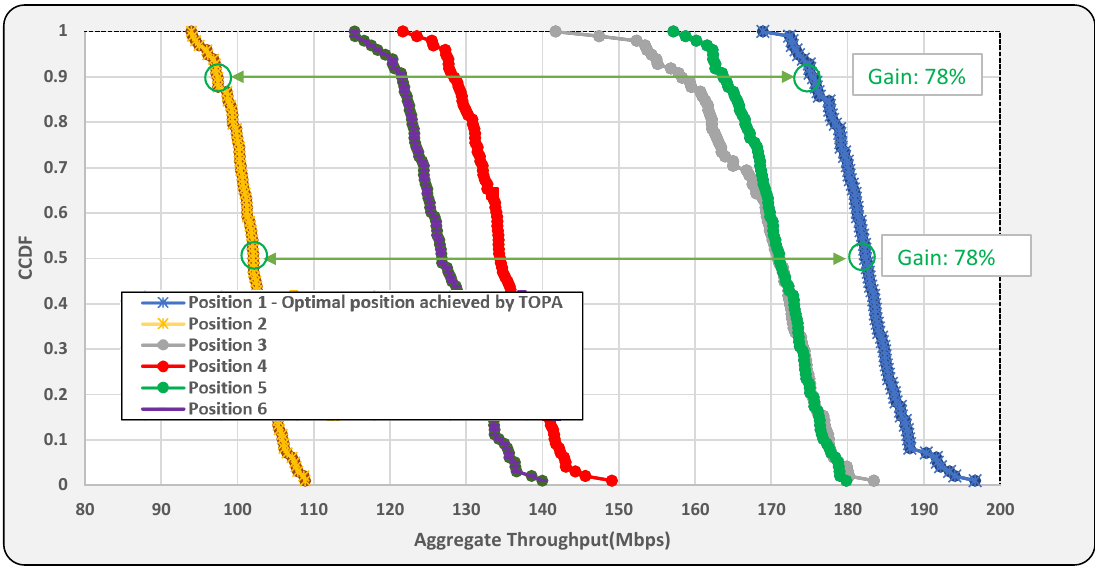}
	\caption{Scenario B - Aggregate throughput measured on UAV, where $\lambda_1 ~\smash{=} ~2 \times \lambda_2 ~\smash{=} ~4 \times \lambda_3 ~\smash{=} ~8 \times \lambda_4$.}
	\label{fig8}
\end{figure}
 The performance achieved in scenario B is depicted in \cref{fig6}, considering all users with the same traffic demand. The aggregate throughput measured on the UAV shows a significant improvement when the UAV is deployed at the optimal position (Position 1). In comparison to Position 2, the aggregate throughput is increased up to 95\% for the 90th percentile and 100\% for the 50th percentile. The other four positions have similar performance due to the similarity of their traffic demand, and the fact that each position allows for LoS connectivity with three groups of UEs. In \cref{fig7}, the traffic demands of $UE_1$ and $UE_2$ are twice the other UEs. In that case, the aggregate throughput is enhanced up to 94\% for the 90th percentile and 80\% for the 50th percentile when compared with Position 2. Finally, for traffic demand $\lambda_1 = 2 \times \lambda_2 = 4 \times \lambda_3 = 8 \times \lambda_4$, there is an increase of 78\% in the aggregate throughput (illustrated in \cref{fig8}). As a result, deploying the UAV at the optimal position results in substantial improvements in aggregate throughput, ranging from 80\% to 100\% across various scenarios. The performance is notably influenced by factors like traffic demand and Line-of-Sight (LoS) connectivity with the users.\looseness=-1

\section{Conclusions~\label{sec:conclusions}}
This work aimed at positioning a UAV in a challenging environment, considering the obstacles in the scenario to establish LoS connections between the UAV and all users. TOPA demonstrates its effectiveness in improving the performance of FNs by optimally positioning the UAV based on traffic demand and ensuring LoS connections. TOPA achieves significant enhancements in aggregate throughput compared to the baseline positions. The results show that the optimal placement of the UAV leads to throughput improvements ranging from 80\% to 100\%. Overall, the findings highlight the potential of TOPA to optimize aerial networks and provide valuable insights for future research in this domain. For instance, the use of computer vision techniques to detect the position of users and obstacles, as well as the evolution to multi-UAV and multi-obstacle scenarios, can be interesting topics for future works.\looseness=-1

\bibliographystyle{IEEEtran}
\bibliography{IEEEabrv,References}

\end{document}